\documentclass[fleqn,usenatbib]{mnras}

\usepackage[T2A,T1]{fontenc}
\usepackage[utf8]{inputenc}
\usepackage{ae,aecompl}

\usepackage{graphicx}	% Including figure files
\usepackage{amsmath}	% Advanced maths commands
\usepackage{amssymb}	% Extra maths symbols

\title[Black hole spin]{Black hole spin from wobbling and rotation of the M87 jet and a sign of a magnetically arrested disc}

\author[D. N. Sob'yanin]{
Denis Nikolaevich Sob'yanin
\fontencoding{T2A}\selectfont
 (Денис Николаевич Собьянин)\thanks{E-mail: sobyanin@lpi.ru}
\fontencoding{T1}\selectfont
\\
I. E. Tamm Division of Theoretical Physics, P. N. Lebedev Physical Institute of the Russian Academy of Sciences,\\Leninskii Prospekt 53, Moscow 119991, Russia
\\
Moscow Institute of Physics and Technology (State University), Institutskii Pereulok 9, Dolgoprudnyi 141701, Russia
}

\date{Accepted 2018 May 25. Received 2018 May 22; in original form 2018 May 11}

\pubyear{2018}
\volume{479}
\pagerange{L65--L69}

\begin{document}
\label{firstpage}
\maketitle

\begin{abstract}
New long-term Very Long Baseline Array observations of the well-known jet in the M87 radio galaxy at 43~GHz show that the jet experiences a sideways shift with an approximately 8--10 yr quasi-periodicity. Such jet wobbling can be indicative of a relativistic Lense-Thirring precession resulting from a tilted accretion disc. The wobbling period together with up-to-date kinematic data on jet rotation opens up the possibility for estimating angular momentum of the central supermassive black hole. In the case of a test-particle precession, the specific angular momentum is $J/Mc=(2.7\pm1.5)\times10^{14}$~cm, implying moderate dimensionless spin parameters $a=0.5\pm0.3$ and $0.31\pm0.17$ for controversial gas-dynamic and stellar-dynamic black hole masses. However, in the case of a solid-body-like precession, the spin parameter is much smaller for both masses, $0.15\pm0.05$. Rejecting this value on the basis of other independent spin estimations requires the existence of a magnetically arrested disc in M87.
\end{abstract}


%%%%%%%%%%%%%%%%%%%%%%%%%%%%%%%%%%%%%%%%%%%%%%%%%%

%%%%%%%%%%%%%%%%% BODY OF PAPER %%%%%%%%%%%%%%%%%%
\section{Introduction}

One of the most well-known and well-studied extragalactic jets, the jet in the giant elliptical galaxy Messier 87 (M87, NGC 4486, 3C 274, Virgo A), occupies a special place among relativistic jets in active galactic nuclei (AGNs). Discovered 100~yr ago \citep{Curtis1918}, the jet still remains one of the main targets of modern theoretical and observational research. One of the nearest, with a distance of only 16--17~Mpc \citep{BlakesleeEtal2009,BirdEtal2010}, and simultaneously having as an AGN central engine a central supermassive black hole with a large mass of $(3-6)\times10^9\text{ }M_\odot$ \citep{GebhardtEtal2011,WalshEtal2013} thus implying a large Schwarzschild radius of about $(1-2)\times10^{15}$~cm, it is the most promising cosmic laboratory site on which one can try to directly discern the seminal processes near the very central engine that give life to AGN jets.

Multifarious emission from the M87 jet is observed throughout the spectrum from radio to TeV \citep{WilsonYang2002,AcciariEtal2009,PerlmanEtal2011,AvachatEtal2016,WongEtal2017}, and radio band is especially important as Very Long Baseline Interferometry (VLBI) thanks to very high, particularly at mm-wavelengths, angular resolution approaching 6--10 Schwarzschild radii \citep{HadaEtal2016,KimM87Etal2016} is an appropriate tool for imaging the fine structure of the jet and observing kinematics of the relativistic flows. In simultaneous observations of the jet in different bands, radio allows one to more precisely localize phenomena (say, flares) seen in other bands \citep{AcciariEtal2009,AbramowskiEtal2012,HadaEtal2014}.

VLBI imaging uncovered key features of the M87 jet, such as apparent superluminal proper motions, limb brightness, wide opening angle at the base, possible recollimation features, parabolic structure relatively near the core and its subsequent transition to a conical shape, and the existence of a counter-jet \citep{JunorEtal1999,LyEtal2004,KovalevEtal2007,AsadaNakamura2012}. Later, the studies of the jet kinematics at scales from 100 to 1000 Schwarzschild radii from the core resulted in detection of the jet rotation and its measurement \citep{MertensEtal2016}. Subsequently, not the simple limb-brightening but a persistent triple-ridge structure across the jet was resolved at 1.6 and 5~GHz \citep{AsadaEtal2016} and 15~GHz \citep{Hada2017}. These new observations allowed us to conclude that this profile structure reflects the intrinsic structure of the jet such that we have not a single jet but jet in jet \citep{Sobyanin2017}.

Recently, \citet{WalkerEtal2018} have presented observational results from the 2007--2008 program of intensive monitoring of the M87 jet at parsec and subparsec scales together with roughly annual observations from 1999 to 2016 using the Very Long Baseline Array (VLBA) radio data at 43~GHz (7~mm) with a resolution of about 30 by 60 Schwarzschild radii. Among other jet characteristics, the authors declare detection of a curious sideways shift of the jet with an approximately 8--10~yr quasi-periodicity that was unobserved before. The aim of this paper is to consider a new way of extracting the spin parameter of the central supermassive black hole in M87 from the known rotational characteristics of the M87 jet and from this new observational phenomenon of jet wobbling.

\section{Jet properties}

\subsection{Rotation}

Here we first describe for clarity what physical laws provide the way for determination of the rotation frequency at the jet base from the observations of the jet dynamics relatively far from the central engine and then give the inferred rotation frequency and launching radius for the M87 jet.

The jet is governed by relativistic magnetohydrodynamics (MHD), and for an ideal plasma we have the Maxwell equations, infinite conductivity condition, and laws of conservation of matter, energy, and momentum, together with an equation of state. If we consider the stationary and axisymmetric case, we have a number of integrals of motion conserved along the magnetic tubes, including the magnetic flux, the quantities reflecting, respectively, matter, energy, and momentum conservation,
\begin{gather}
\label{GSeta}
\eta=\frac{\gamma\rho v_\text{p}}{B_\text{p}},\\
\label{GSE}
\mathcal{E}=\gamma h\eta-\frac{\Omega_\text{F}I}{2\pi},\\
\label{GSL}
\mathcal{L}=\gamma h\eta r v_\phi-\frac{I}{2\pi},
\end{gather}
and the quantity reflecting infinite conductivity and named the Ferraro isorotation frequency,
\begin{equation}
\label{GSOmegaF}
\Omega_\text{F}=\frac{v_\phi-v_\text{p}B_\phi/B_\text{p}}{r},
\end{equation}
where $v_\phi=\Omega r$ and $B_\phi$ are the toroidal and $v_\text{p}$ and $B_\text{p}$ are the poloidal components of the velocity and magnetic field,  $\gamma$ is the Lorentz factor, $h=1+\varepsilon+p/\rho$ is the specific relativistic enthalpy, $\varepsilon$ is the specific internal energy, $p$ is the pressure, $\rho$ is the density in the comoving frame, and $I$ is the electric current in the magnetic tube ($G=c=1$ throughout the paper). These integrals allow one to take into account the changing of the jet radius with distance from the base and, if found at some distance, can be used to calculate various physical quantities at other distances from the base.

The isorotation frequency \eqref{GSOmegaF} equals the actual angular frequency at the jet base, where $v_\text{p}$ vanishes. Being an integral of motion, it can be estimated from the actual angular frequency, radius, and Lorentz factor of the jet taken at some distance from the base. Such an estimation is based on conservation of the quantity
\begin{equation}
\label{l}
l=\gamma h(1-\Omega_\text{F}rv_\phi)
\end{equation}
along the magnetic tube, which follows from a combination of the integrals \eqref{GSeta}--\eqref{GSL}. Putting $h\approx1$ and $l\approx1$, the former reflecting a cold flow and the latter corresponding to nonrelativistic motion at the base, we arrive at
\begin{equation}
\label{OmegaFEstimation}
\Omega_\text{F}\approx\frac{1-\gamma^{-1}}{\Omega r^2}.
\end{equation}

With the wavelet-based image segmentation and evaluation method applied to the VLBA data, \citet{MertensEtal2016} have studied kinematics of the M87 jet on the linear scales down to 100 Schwarzschild radii. The authors analyse two regions in the VLBA 7-mm radio images of the jet jointly covering the range 0.5--4 marcsec from the core and under the two conditions mentioned above (nonrelativistic motion at the base and a cold flow) estimate from the inferred 2D flow kinematics the isorotation frequency
\begin{equation}
\label{M87OmegaF}
\Omega_\text{F}=(1.1\pm0.3)\times10^{-6}\text{ s}^{-1}
\end{equation}
and the corresponding Keplerian jet-launching radius
\begin{equation}
\label{M87RBase}
R_\text{base}=(4.8\pm0.8)R_\text{Sch},
\end{equation}
where $R_\text{Sch}=2 M$ is the Schwarzschild radius of the black hole and $M$ is its mass. These quantities will be used below when estimating the black hole spin.

\subsection{Wobbling}

\citet{WalkerEtal2018} reported new results of the M87 VLBA observational program at 43~GHz. The program was initially devoted to fast sampling of the processes near the M87 core aiming at determination of apparent superluminal motions in the jet. Meanwhile, attempts to find radio counterparts to TeV flares in M87 started from 2009 together with earlier archival data resulted in roughly annual observations of the jet over the 17-yr period starting from 1999. This allowed one to trace the long-term dynamics of the jet and, as a significant byproduct, to find that the jet moves transversely on time-scales of several years.

Specifically, \citet{WalkerEtal2018} have found significant transverse displacement of the jet, especially in the range 2--8 marcsec from the core. The overall displacement dynamics is consistent with a gradual linear change in the jet position angle and an extra quasi-sinusoidal variation. Modelling the data with an empirical equation not based on any physical model gives a period for the variation of $10.3\pm0.3$~yr. At the same time, the authors note that the data taken solely from 2 and 3~marcsec from the core imply a different period result of $7.6\pm0.3$~yr. The latter data have a special place in the sense that these are more lengthy and include additional 7~yr of observations before 2006, while the remaining data cover the range only from 2006 to 2016. Thus, the authors report an approximate 8--10~yr quasi-periodicity of the sideways shift.

The situation may be clarified after several extra years of high-quality observations covering, say, one more entire period of such jet `wobbling'. In view of the present uncertainty, we have to adopt a rough wobbling period of
\begin{equation}
\label{Twobbling}
T_\text{wob}=9\pm1\text{ yr}.
\end{equation}

\section{Black hole spin}

How can the described wobbling of the M87 jet be explained? \citet{WalkerEtal2018} mention that the wobbling is a natural consequence of the jet acceleration and collimation process, as modern 3D general relativistic MHD (3D GRMHD) simulations show \citep{TchekhovskoyEtal2011}, and may reflect a Kelvin-Helmholtz instability in a jet with a density not exceeding that in the ambient environment \citep{Hardee2007}. Note that MHD is an approximation with its limits of applicability and is less comprehensive than kinetic theory, so in order to catch, say, kinetic jet instabilities, one has to resort to particle-in-cell simulations \citep{NishikawaEtal2017}.

At the same time, \citet{LiskaEtal2018} have conducted most recent high-resolution 3D GRMHD simulations of tilted accretion discs around rotating black holes. Numerical simulations of black hole accretion flows have a long history and were carried out in HD and MHD setting first for the case of alignment of the disc and black hole equatorial planes \citep{Wilson1972,Hawley1991,KoideEtal1999,GammieEtal2003} and subsequently for the case of misalignment \citep{FragileAnninos2005,FragileEtal2007}. There also appeared simulations of not only accretion per se but also the concomitant production of jets \citep{HawleyKrolik2006,McKinneyNarayan2007,TzeferacosEtal2009,PorthEtal2011,MoscibrodzkaEtal2016}. Against this background, \citet{LiskaEtal2018} have found that discs that are tilted can produce magnetized relativistic jets, which propagate along the disc rotation axis, not along the black hole rotation axis. In addition, the produced jets undergo the Lense-Thirring precession \citep{Thirring1918,LenseThirring1918,Thirring1921} together with the disc.

\subsection{Lense-Thirring precession}

Let us consider the possibility that the observed wobbling of the M87 jet reflects the Lense-Thirring precession. This type of precession appears due to the GR frame-dragging effect when the orbit of a test particle is tilted with respect to the equatorial plane of a rotating black hole. The angular frequency of the Lense-Thirring precession is \citep{Wilkins1972}
\begin{equation}
\label{OmegaLT}
\Omega_\text{LT}=\frac{2J}{R^3},
\end{equation}
where $J$ is the angular momentum of the black hole and $R\gg M$ is the radius of the orbit. Using the Keplerian velocity distribution $v=\sqrt{M/R}$ under the same condition on radius allows us to work directly with the Ferraro isorotation frequency \eqref{M87OmegaF} instead of the Keplerian jet-launching radius~\eqref{M87RBase} and to write out the specific angular momentum of the black hole, having dimensions of length,
\begin{equation}
\label{JOverM}
\frac{J}{M}=\frac{\Omega_\text{LT}}{2\Omega_\text{F}^2}.
\end{equation}
This manoeuvre alleviates the existing factor-of-two uncertainty in determining the mass of the central supermassive black hole in M87. We then get from \eqref{M87OmegaF} and \eqref{Twobbling}
\begin{equation}
\label{M87JOverM}
\frac{J}{M}=(2.7\pm1.5)\times10^{14}\text{ cm}.
\end{equation}

In order to estimate the dimensionless spin parameter
\begin{equation}
\label{a}
a=\frac{J}{M^2},
\end{equation}
which represents the specific angular momentum in units of gravitational radii $R_\text{Sch}/2$, we, however, have to consider different black hole masses. The mass estimated from gas dynamics \citep{WalshEtal2013},
\begin{equation}
\label{BHMassGD}
M_\text{GD}=3.5^{+0.9}_{-0.7}\times10^9 M_\odot,
\end{equation}
implies the spin parameter
\begin{equation}
\label{aGD}
a_\text{GD}=0.5\pm0.3.
\end{equation}
In turn, the mass estimated from stellar dynamics is almost twice as large  \citep{GebhardtEtal2011},
\begin{equation}
\label{BHMassSD}
M_\text{SD}=(6.0\pm0.4)\times10^9 M_\odot
\end{equation}
[rescaled for a distance to M87 of 16.4~Mpc \citep{BirdEtal2010}], so the spin parameter is correspondingly smaller,
\begin{equation}
\label{aSD}
a_\text{SD}=0.31\pm0.17.
\end{equation}
Large spin uncertainties are determined mainly by uncertainties in $\Omega_\text{F}$ stemming from difficulties of extracting the exact jet kinematics from VLBI observations, the latter uncertainties doubling because of the square $\Omega_\text{F}$ dependence entering~\eqref{JOverM}.

Note that relations \eqref{JOverM} and \eqref{a} have no structural form or restrictions always leading to an inferred spin below unity irrespective of the values of the wobbling period and Ferraro isorotation frequency, so one could potentially get, say, $10^{-3}$ or 1000. Otherwise one could substitute the periods of various oscillating phenomena and always obtain reasonable spin values, whether or not these are related to the Lense-Thirring precession. Thus, if we even abstract ourselves from the exact values and uncertainties of the inferred black hole spin, the sole circumstance that the obtained spin values do not exceed unity and are not very small favours that the observed wobbling of the jet can indeed result from the Lense-Thirring precession.

\subsection{Solid disc}

We have just considered the case of a test-particle Lense-Thirring precession, and now let us see what changes when the accretion disc precesses as a solid body. This situation takes place when the sound-crossing time for the disc is short compared with the precession time \citep{FragileEtal2007}. The Lense-Thirring angular frequency for a global solid-body-like precession is \citep{IngramEtal2009}
\begin{equation}
\label{OmegaLTsolid}
\Omega_\text{LT}^\text{solid}=10 J \frac{r^{-1/2}-R^{-1/2}}{R^{5/2}-r^{5/2}}
\end{equation}
and corresponds to averaging the value \eqref{OmegaLT} over the whole disc, having a constant surface density, from the inner radius~$r$, taken as the radius $r_\text{ISCO}$ of the innermost stable circular orbit (ISCO), to the outer radius~$R$, taken as the jet launching radius~\eqref{M87RBase}, so that the frequency now has an extra dependence on spin parameter via inner radius. More complex relations taking into account the spin dependence of the local Lense-Thirring and Keplerian frequencies can be found in \citet{FranchiniEtal2016,MottaEtal2018} with useful polynomial approximations for numerical simulations in \citet{FalcoMotta2018}.

It follows from \eqref{a} and \eqref{OmegaLTsolid} that the dimensionless spin parameter is defined implicitly via the relation
\begin{equation}
\label{aImplicit}
a=\frac{\pi M}{5T_\text{wob}}\frac{(M\Omega_\text{F})^{-5/3}-(r_\text{ISCO}/M)^{5/2}}{(r_\text{ISCO}/M)^{-1/2}-(M\Omega_\text{F})^{1/3}},
\end{equation}
where $r_\text{ISCO}/M$ is the ISCO radius measured in units of gravitational radii and dependent on~$a$ \citep{BardeenEtal1972}. Several iterations starting from the test-particle spin values found above give the spins in the case of a solid-body-like disc precession for the gas-dynamic and stellar-dynamic black hole masses, respectively,
\begin{gather}
\label{aGDsolid}
a_\text{GD}^\text{solid}=0.16\pm0.05,\\
\label{aSDsolid}
a_\text{SD}^\text{solid}=0.14\pm0.04.
\end{gather}
Due to proximity of the values we may adopt a single spin parameter for a solid-body-like disc precession,
\begin{equation}
\label{aSolid}
a_\text{solid}=0.15\pm0.05.
\end{equation}

The spin obtained for the case of a solid-body-like precession is significantly lower than that for the case of a test-particle precession. This is natural as in the former case all inner orbits from jet-launching radius down to ISCO that have higher local Lense-Thirring frequencies contribute to the global precessional motion of the disc, so that the common frequency is higher for the same spin and, correspondingly, the same precession period requires a lower spin.

\section{Discussion}

Let us consider previous spin estimations for the M87 black hole. Using the observed rapid TeV variability in M87 and estimating the optical depth of the radiation field from an advection-dominated accretion flow (ADAF) to TeV photons, \citet{WangEtal2008} proposed $a>0.65$. The physical principle of such an estimation is that, in order to be visible, TeV photons should escape from the innermost regions of the disc, where these are presumably formed, through the ADAF radiation fields, dependent on spin. Then, \citet{LiEtal2009} extended the previous model based on self-similar ADAF solutions in Newtonian approximation and considered GR effects for a HD radiatively inefficient accretion flow (RIAF), which gave $a>0.8$. Numerically modelling spectral fits to the M87 core data from radio to hard X-ray under the condition that all the emission goes from the immediate surroundings of the central black hole, \citet{HilburnLiang2012} offered the same constraint from the best-fitting parameters. At the same time, \citet{DoelemanEtal2012} carried out 1.3-mm VLBI observations of the M87 core and found a full width at half-maximum size of $(5.5\pm0.4)R_\text{Sch}$. Assuming it as the ISCO diameter and taking into account its $a$ dependence, the authors concluded that the disc orbits in a prograde sense and that $a>0.2$. Recently, \citet{FengWu2017} interpreted a mm-bump found by \citet{PrietoEtal2016} in high-resolution multiwaveband observations of M87 as synchrotron emission of thermal electrons in the ADAF and, with constraints on accretion rate and using a model jet-power dependence, estimated a spin of $a=0.98$.

These spin estimations are model dependent and based on assumptions about ADAFs/RIAFs and concomitant radiation spectra, implicit fiducial parameters and relations in the disc and jet models, or observational parameters known to a factor of a few, such as the M87 jet power. The most direct estimation utilizes only the core size but is based on the assumption about the ISCO determining that size \citep{DoelemanEtal2012}. The estimations presented in this paper are also direct and utilize the jet rotation and wobbling only, the first giving the jet-launching size and the second the precession frequency. In a sense, the extra precession measurement is the cost we should pay for removing the assumption about the relation of the ISCO and jet-launching radii. Another cost is, however, that the estimations depend on the mass distribution in the accretion disc in the case of a solid-body-like precession. On the other hand, this can allow us to draw some conclusions about the disc structure if we use an independent spin estimation. Let us stress that estimating the black hole spin for M87 has a power in constraining general models of jet formation as certain models of discs and jets prefer certain black hole spin values \citep{TchekhovskoyEtal2011}.

The values of the black hole spin in M87 obtained in this paper can be considered moderate or even low. Generally, the low spin values cannot be rejected. \citet{Reynolds2014} have analysed the existing data on spins of supermassive black holes in AGNs obtained with X-ray reflection spectroscopy and found a probable trend of decreasing spin with increasing mass, so that low spins are not forbidden for rotating black holes with mass comparable to that of the M87 black hole and may be even more natural. Radio-loudness of M87 also does not deny moderate or small spin values because the black hole spin itself cannot allow one to distinguish between the radio-loud and radio-quiet types of the AGN \citep{Reynolds2014,GarofaloEtal2014}.

An afterthought is that if the inferred spin is probably too low for a solid-body-like precession of a constant surface density disc, then we do not have enough medium at the base inside the jet-launching radius to provide the necessary surface density and Lense-Thirring frequency. Interestingly, the developed jet-in-jet model for M87 \citep{Sobyanin2017} already has implicit indications of this. The jet as a whole in fact consists of two coaxial embedded jets such that the outer jet is an annular hollow plasma cylinder that contains a narrow inner jet. The rotating relativistic inner and outer jets gradually widen with distance and are separated by an interlayer of a low-density plasma with electromagnetic fields. The low density in the interlayer can reflect the low density at the base inside the jet-launching radius.

This relativistic ideal MHD model allowed us to find various specific physical quantities for the M87 jet, including electromagnetic fields, charges and currents, pressures, densities, multiplicities, mass fluxes, and temperature. Particularly, the total mass flux through the jet is determined mainly by the mass flux in the outer jet and equals
\begin{equation}
\label{M87totalMassFlux}
\dot{M}\sim0.05\text{ M}_\odot\text{ yr}^{-1}.
\end{equation}
Importantly, this value is very large and comparable to the measured Bondi accretion rate $0.1-0.2\text{ M}_\odot\text{ yr}^{-1}$ across the Bondi radius of M87 at $0.12-0.22$~kpc \citep{RussellEtal2015} (note that the flux \eqref{M87totalMassFlux} should be doubled because of the existing counter-jet). This circumstance favours a scenario when almost all initial accretion flow far from the jet goes to the outer jet so that the jet can substantially suppress accretion on to the black hole.

This situation corresponds to the so-called magnetically arrested disc (MAD) \citep{NarayanEtal2003,Igumenshchev2008,TchekhovskoyEtal2011}, so that the flow is stopped at the jet-launching radius by the magnetic field and then transmitted to the jets while the accretion is impeded. The balance of the magnetic and ram pressures at the base determines the density responsible for the mass flux \eqref{M87totalMassFlux} in the outer jet \citep{Sobyanin2017}. The fact that the black hole spin parameter for a solid-body-like accretion is low and is less than the estimation of \citet{DoelemanEtal2012} may be an extra indication of the MAD in M87. The spin values for a test-particle precession are already higher and correspond to the situation when we have moved the medium between the ISCO and jet-launching radii to the latter radius, thus forming a dense ring. In fact, we probably have a solid-body-like precession of the disc the inner radius of which is not the ISCO radius but the jet-launching radius and the outer radius of which is not the jet-launching radius but some characteristic radius $R_\text{acc}$ reflecting the scale of accumulation of the matter outside the jet due to the stopping effect of the magnetic field. Then using \eqref{OmegaLTsolid} under the described conditions and adopting $a=0.98$ of \citet{FengWu2017} (result unchanged for the Thorne limit $0.998$ or formal maximum $1$) gives for $R_\text{base}\approx5R_\text{Sch}$ for the stellar-dynamic and $R_\text{base}\approx7R_\text{Sch}$ for the gas-dynamic black-hole mass almost the same matter accumulation radius in the MAD,
\begin{equation}
R_\text{acc}\approx10 R_\text{Sch}.
\end{equation}

Incidentally, if we look at the problem from another side and assume a MAD solely from observing jet in jet and from a large mass flux, then the obtained spin values for a test-particle precession give the lower bound for the M87 black hole spin, and this may be an independent sign of the MAD preferring moderate or high spins and disliking low spins when generating efficient outflows. This is consistent with the present simulations \citep{TchekhovskoyEtal2011,McKinneyEtal2012}.

Note finally that the observed jet wobbling, if due to the Lense-Thirring precession, can be considered, in addition to the entire jet-in-jet structure and $R_\text{base}$ estimations, evidence for the Blandford-Payne mechanism launching the outer jet in M87.

\section{Conclusions}

In this paper, long-term VLBA observations of the M87 jet at 43 GHz are considered that show a new phenomenon of quasi-periodic jet wobbling. A new independent way to estimate the spin of the central supermassive black hole in M87 is proposed, which uses the wobbling period and kinematic data on jet rotation and is based on the assumption that the nature of the observed wobbling is a relativistic Lense-Thirring precession of a tilted accretion disc. A test-particle precession gives medium spins, while a solid-body-like precession gives low spins. The low values, if rejected on the basis of other independent spin estimations, can indicate that the accretion disc cannot continue down to the ISCO radius and is seemingly a MAD.

\section*{Acknowledgements}

The work is supported by the Russian Science Foundation, grant no. 16-12-10051.

%%%%%%%%%%%%%%%%%%%%%%%%%%%%%%%%%%%%%%%%%%%%%%%%%%



\bibliographystyle{mnras}
\providecommand{\noopsort}[1]{}\providecommand{\singleletter}[1]{#1}%

%%%%%%%%%%%%%%%%%%%%%%%%%%%%%%%%%%%%%%%%%%%%%%%%%%

% Don't change these lines
\bsp	% typesetting comment
\label{lastpage}
\end{document}